\documentclass[12pt]{article}

\usepackage[utf8]{inputenc}
\usepackage[english]{babel}

\usepackage{listings}
\usepackage{courier} 

\usepackage[round]{natbib}
\usepackage{hyperref}

\usepackage[pdftex]{graphicx}

\usepackage{pgf}
\usepgflibrary{arrows.meta}
\usepackage{rotating}

\usepackage{amsmath}
\usepackage{amssymb}
\usepackage{amsthm}

\usepackage[figure,table]{hypcap}

\hypersetup{
    colorlinks,%
    citecolor=blue,%
    filecolor=blue,%
    linkcolor=blue,%
    urlcolor=blue
}

\lstdefinelanguage{tscore}{basicstyle=\ttfamily\scriptsize,
xleftmargin=2em,
aboveskip=\smallskipamount,belowskip=\smallskipamount
,moredelim=[is][\metabox]{/$}{$/},
     numbers=left,
     numbersep=1pt,
     numberstyle=\tiny,
     emph={T,VOX,PARS},emphstyle=\textbf
}

\title{
  Data Models of German Lute Tablature\\With \Tscore
}
\author{Markus Lepper and Baltasar Trancón y Widemann}
\date{(20231222--20240524)}

\newcounter{xxnum}

\newcommand{\xsimpl}{\ensuremath{\sim\kern-1.0ex+} }
\newcommand{\xsimst}{\ensuremath{\sim\kern-1.0ex*} }
\newcommand{\xumbi}[1]{#1}

\newcommand{\xxt}[1]{``\texttt{#1}''}

\newcommand{\makeobjectTT}[1]%
{\expandafter\newcommand\csname x#1\endcsname{\ensuremath{\mathtt{#1}}}}
\newcommand{\makeobjectIT}[1]%
{\expandafter\newcommand\csname x#1\endcsname{\ensuremath{\mathit{#1}}}}
\newcommand{\makeobjectSF}[1]%
{\expandafter\newcommand\csname x#1\endcsname{\ensuremath{\mathsf{#1}}}}
\newcommand{\makeobjectBF}[1]%
{\expandafter\newcommand\csname x#1\endcsname{\ensuremath{\mathbf{#1}}}}

\newcommand{\tscore}{\textsf{tscore}}
\newcommand{\Tscore}{\textsf{Tscore}}

\begin{document}
%
\maketitle
\begin{abstract}
\Tscore{} is both an abstract formalism and its computer implementation
to construct models of arbitrary kinds of time-related data.
It is a research project about the semantics of musical notation,
applying the method of computer-aided re-modelling to diverse
formalisms and semantics of time-related data.
Here we present the application to German tablature notation.
While the current implemention is merely a proof of concept, the
lean architecture of \tscore{} allows easy adaptation and extension.
\end{abstract}
%

\section{Music and Computer Usage}

Beginning with approximately 1970, the use of
  computers, up to then only affordable by big institutions, became 
  accessible for everbody's personal work.
  Soon first applications in the realm of music execution, processing, notation,
  etc.\ came up.
  In the last twenty years, the processing power of affordable
  hard- and software increased to make also complex applications feasible
  for everyone.

  Nowadays all arts and sciences dealing professionally
  with historical matters can
  (and sometimes even: must) apply computer technology for their daily business.
  For instance, the realm of \emph{digital humanities} comprises the art of
  encoding, storing, archiving, analyzing even antique and medieval text corpora
  with computer technology. This indeed allows in any case much faster and
  lossless retrieval,
  wider public access, and more convenient transport and storage.
  Even more: sometimes it enables fundamentally new techniques of comparing,
  annotating, retrieving, and analyzing.

  There are three main axes which can guide the selection of the digital concepts
  and products to apply:

  \textbf{Commercial vs.\ academic providers:}\\
  Of course the whole realm of digital humanities is commercially not as
  profitable as, for instance, gaming, logistics, transport, or entertainment.
  Nevertheless also here two large groups of providers can be distinguished:

  Commercial vendors want to sell particular products
  (computer program, storage space, consulting, etc.) 
  Their advantage is that in most cases you can expect stable
  sales structures over the next years, including legal obligations for
  service and repair.

  Acadamic players are interested more in the research aspect and supply the
  computer software for free.
  That legal obligations are missing, but normally compensated or even
  over-compensates by making the source code public, and by a vivid community
  of users and volunteer developers.

  \textbf{Closed vs.\ open systems:}\\
  This distinction often, but not always, runs in parallel with the first.

  Closed systems support fixed and (more or less precisely) defined input/output
  formats.
  The internal data is neither visible nor accessible, its structure is not known,
  the possible manipulations are only those foreseen by the vendor.

  An open system contrarily comes with a public definition of its
  data model and with interfaces to retrieve and manipulate this model
  by user-controlled program code.
  
  Somehow in between stand the programs which are not fully transparent, but 
  which allow to import and export an XML representation
  of their internal data. So the data can be subject to all transformations
  applicable to this wide-spread standard data format.

  Both groups have their advantages:
  
  With a closed system you can for instance create a printable and readable
  music notation rendering by using the interactive ``clicks and drags'' foreseen
  by the vendor. This can be the more convenient way for a user not
  familiar with computer programming.
  But with an open system a skilled programmer can generate data models
  programmatically, and freely add user-defined functions for analyzing,
  transforming, rendering etc.

  Examples for the closed systems are ``Microsoft Word'' and ``Finale'', versus
  the open ``LaTeX'' and ``lilyPond''.

  \textbf{Interactive vs.\ text input:}\\
  A third axis for categorization is the difference between user actions
  and text files as the (predominant) means for input.
  Interactive systems let the user construct the data model by executing
  activities, in most cases on some graphical user interface (GUI). In case of
  music input also MIDI keyboards and other controllers can be supported.
  The main advantage is the early achievement of first simple
  results (``shallow learning curve'').

  The compiler-like approach lets the user write a source text which is then 
  (as a digital text object, constructed with any text editing program)
  translated into the intended data model.
  Text-based systems are more appropriate to larger corpora, because they allow 
  to apply all generic text processing techniques like searching, replacing,
  spell-checking, comparing, sorting, version control, automated
  analysis and translation, etc.
  Their critical data can be read, written, and discussed with only pen and
  paper, and often their semantics are defined in a much more precise way. 

  Significant examples are again   ``Microsoft Word'' and ``Finale'',
  versus ``LaTeX'' and ``lilyPond''.\footnote{
  Be warned that systems which employ coding standards like XML or JSON as
  the basis for their data representation can also be called text-based, but merely
  in a technical sense: Both formats are hardly writeable, readable, and
  maintainable by humans.
  }

\section{The Context of \Tscore}

\Tscore{} is both an abstract formalism and its computer implementation
to construct models of arbitrary kinds of time-related data.
It is in the public domain, open source, implements an open data model,
and follows the text-based compiler approach.
The data models are always well-specified by construction and
accessible by user-defined Java code, if required.
\Tscore{} has been successfully employed for very diverse kinds of such data,
like abstract films, conventional music notation, avantgarde graphic notation,
sound synthesis parameters, etc. \citep{tscoreSW,keod2013,translets2016}

The complete picture is an architecture of three layers:
\Tscore{} is a meta-meta-model, which allows
by plugging-in and wiring particular definitions of parsers, data types, and
storages (by writing snippets of Java code)
to define meta-models, the instances of which are the models of the scores,
opera, or other aesthetic objects.

\Tscore{} as a research project deals with the \emph{semantics of notation}:
By experimentally re-modelling the semantics of a notation system (= by realizing
this meta-model by a computer application), its limitations and implications on the
possible models are made visible.
In this context it is crucial that \tscore{} does not
model a particular syntax as such, but its meaning (= its semantics) in a
much more direct way.

The high degree of organization of the \tscore{} core and libraries allows to define
meta-models and models with minimal effort: The implementation of the complete
processing pipeline described
below needs only about 500 lines of code,\footnote{
480 lines of Java and 30 of DTD, stripped from comments, counted using \xxt{cloc}.
}
and entering the complete model of \emph{Die Kunst der Fuge}
required not more than four working days.\footnote{
The voices of the canons and the scores of the mirror fugues were only entered once.
}

Being experimental research, the code, the underlying mathematics, and the
structural insights
are enhanced by applying \tscore{} in a practical way progressively to
new realms. A presentation
of the ``E-laute'' project about German Lute Tablature \citep{lute} on the
``Encoding Cultures
--- Joint MEC and TEI Conference Paderborn 2023'' inspired us to apply
\tscore{} to this topic.

The aim is to implement a processor which allows easy construction of data models
which can be subject to very diverse processing. The input format shall
resemble the historic original as much as possible, shall be readable by both
computers and humans, and writeable even with pen and paper.

The \tscore{} models parsed into a computer can then
be further processed by the exported XML representation,
or by directly inquiring or modifying the model objects by Java language means.

\begin{figure}
{
  \includegraphics[width=0.9\textwidth]{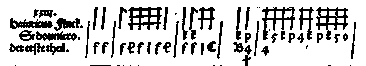}

}\mbox{}\\[1ex]
\begin{lstlisting}[language=tscore]
// Newsidler, nach Wolf, S. 43:

  duratioManet = nonEst
  duratioCadens = nonEst

  Standard_1531_Newsidler_etAlii
   = ( (1 a  f  l  q  x  aa)
       (2 b  g  m  r  y  bb)
       (3 c  h  n  s  z  cc)
       (4 d  i  o  t  &  dd)
       (5 e  k  p  v  C  ee) )

PARS sola

bünde = Standard_1531_Newsidler_etAlii

T       I I T E_ E E _E  I  T F_ _F I I F_ F F _F   F_ _F   E_ E E _E  
VOX v1                   k  k       k p k  5 k p    4  k    p  k 5 o
VOX v2  f f f e  f 1 f   f  f 1  C  & 4 4+ 
    edit                      "hardly readable, could be a '1'"! \\

//eof
\end{lstlisting}
%
%

{\center
  \includegraphics{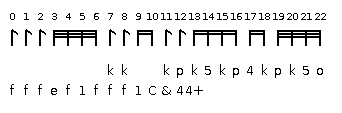}

}
\caption{Example by Newsidler: (1) First line of the original, as reproduced
  by \citet[p.\,43]{wolf};
  (2) \tscore{} source (3) generated control graphic
  \label{fig-newsidler}
}
\end{figure}

\begin{figure}

\begin{lstlisting}[language=tscore]
<tabulatura>
  <columna>
    <duratio source='I' numerus='0' ypos='0' summaPraecedentium.num='0'
          summaPraecedentium.den='1' duratio.num='1' duratio.den='4' />
    <sonum source='f' fret='2' string='0' ypos='2' />
  </columna>
  <columna>
    <duratio source='I' numerus='1' ypos='0' summaPraecedentium.num='1'
          summaPraecedentium.den='4' duratio.num='1' duratio.den='4' />
    <sonum source='f' fret='2' string='0' ypos='2' />
  </columna>
  <columna>
    <duratio source='T' numerus='2' ypos='0' summaPraecedentium.num='1'
          summaPraecedentium.den='2' duratio.num='1' duratio.den='8' />
    <sonum source='f' fret='2' string='0' ypos='2' />
  </columna>
  <columna>
    <duratio source='E_' numerus='3' ypos='0' summaPraecedentium.num='5'
          summaPraecedentium.den='8' duratio.num='1' duratio.den='32' />
    <sonum source='e' fret='1' string='4' ypos='2' />
  </columna>
  <columna>
    <duratio source='E' numerus='4' ypos='0' summaPraecedentium.num='21'
          summaPraecedentium.den='32' duratio.num='1' duratio.den='32' />
    <sonum source='f' fret='2' string='0' ypos='2' />
  </columna>

</tabulatura>
\end{lstlisting}


  \caption{Example by Newsidler: Beginning of the generated XML intermediate
    model, as it appears in XML text files    \label{fig-newsidler-xml}
    }
\end{figure}

\setcounter{xxnum}{1}
\begin{sidewaysfigure}
{
\includegraphics[width=0.9\textwidth]{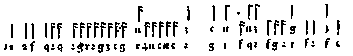}
\includegraphics[width=0.9\textwidth]{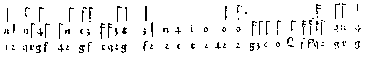}

}\mbox{}\\[1ex]

\begin{lstlisting}[language=tscore]
// Arnolt Schlick, nach Wolf, S. 42/43:

  duratioManet = est
  duratioCadens = est 

  Standard_1531_Newsidler_etAlii
   = ( (1 a  f  l  q  x  aa)
       (2 b  g  m  r  y  bb)
       (3 c  h  n  s  z  cc)
       (4 d  i  o  t  &  dd)
       (5 e  k  p  v  C  ee) )

PARS sola
   bünde = Standard_1531_Newsidler_etAlii

T      I  I  I I  T.  F   F F F F   F F F F    T. F   F F F F   I   I   T.  F   T T  T  F F I I I I I
VOX v1                                         n                z   c   n       n z         g     z  
VOX v2 2  2  f q  2   q   2 g r 2   g z c g    r  4   n c n c   2   g   1   f   q 2  f  g 2 1 f 2 f c


T      I I T T T T   T T T T  F F T   I  I I  - - - - - I T.  F T T T T F F F F T T I
VOX v1 n+  n+  4       n c z      z   c  z    n 4 i o   o o                 z   g n 4
VOX v2 1 2 q r g f   4 2 g f  c q 2   g  f 2  2 c r 2 4 2 2   g z c o C i & q 2 g e g

// eof
\end{lstlisting}



\includegraphics[width=\textwidth]{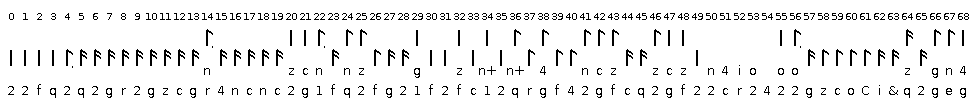}

\caption{Example by Schlick: (1) First two lines
  of the original, as reproduced by \citet[p.\,43]{wolf};
  (2) the corresponding \tscore{} source (3) generated control graphic
  \label{fig-schlick}
}
\end{sidewaysfigure}

\setcounter{xxnum}{1}
\begin{figure}
\begin{lstlisting}[language=tscore]
<!ELEMENT tabulatura (columna)*  >

<!ELEMENT columna (duratio, sonum+) >

<!ELEMENT duratio EMPTY>
<!ATTLIST duratio source CDATA                          #REQUIRED
                  numerus CDATA                         #REQUIRED
                  ypos   (0|1|2|3|4|5|6|7|8|9|10|11|12) #REQUIRED
                  trabes (initialis|terminalis)         #IMPLIED 
                  duratio.num    CDATA                  #REQUIRED
                  duratio.den    (1|2|4|8|16|32|64)     #REQUIRED
                  summaPraecedentium.num  CDATA         #REQUIRED
                  summaPraecedentium.den  (1|2|4|8|16|32|64) #REQUIRED
>

<!ELEMENT sonum EMPTY>
<!ATTLIST sonum source CDATA  #REQUIRED
                  fret   (0|1|2|3|4|5|6|7|8|9|10|11|12) #REQUIRED
                  string (0|1|2|3|4|5|6|7|8|9|10|11|12) #REQUIRED
                  prolongate (yes)                      #IMPLIED
                  ypos   (0|1|2|3|4|5|6|7|8|9|10|11|12) #REQUIRED
                  finger  (p|i|m|a|o)                   #IMPLIED
\end{lstlisting}


\caption{DTD of the (intermediate) result data  \label{fig-dtd} }
\end{figure}

\section{Input Format}

Figure~\ref{fig-newsidler} and \ref{fig-schlick} show the first lines of
two examples given by \citet{wolf} and their translation to the \tscore{} format.

\Tscore{} follows the basic paradigm of conventional Common Western Notation (CWN):
Time flows from left to right over a sheet of paper (or a presentation screen);
synchronously sounding voices are vertically stacked;
one single ideal unlimited horizontal time scale is broken into
staff segments, again vertically stacked.

A \tscore{} source text is a text document edited in any fixed-space font, so that
the text columns in all lines appear aligned. Every source file contains
different sections labelled by the identifier \xxt{PARS}, see
Figure~\ref{fig-newsidler}, line 13.
These sections contain the staves which must start with one time line. This is
lead in by the identifier \xxt{T} and contains the top-level  time-point
organization of this staff (source text line 17).
In CWN meta-models these can be bar numbers and sub-divisions; in electronic scores
this can be plain text representations of seconds and milliseconds.

The German Lute Tablature borrows the duration symbols from late Mensural Notation.
The concrete meaning of the note symbols differs from corpus to corpus, and so
does the relation of these duration symbols to the same symbols appearing
in parallel voice staves---see the Newsidler example \citep[p.\,43]{wolf}.
This semantic layer is not dealt with in the following---we simply encode the
symbols by their graphical appearence, read as if they were  ``modern''
duration symbols, well knowing that this is not a semantic statement but only
a technical encoding.

In the source text we use the capital letters \xxt I, \xxt T,  \xxt F,
and \xxt E,  for stems with zero to three flags,
see Figure~\ref{fig-newsidler} line 17.
The underscore \xxt\_ as a postfix means the start of one
or more beams (replacing all the flags); as a prefix it means the end of
these beams.
(More complex changes of the beam structure, like ending only a subset and
continuing others, cannot be expressed.)
As mentioned above, these durations are encoded
by reading them the modern way, 
thus as $1/4$, $1/8$, $1/16$, and $1/32$.
As a suffix the dot \xxt. has the ``modern'' meaning of multiplying by $3/2$,
see Figure~\ref{fig-newsidler} line~16, and event number~4 in the control graphic.
But standing alone the dots \xxt., \xxt{..}, and \xxt{...} represent the
longer durations, as defined by 
\citet[p.\,41]{wolf}, and are encoded canonically as a $1/2$, $3/4$, and $1/1$.

Below the time lines folllow the voice lines, lead in by the identifier \xxt{VOX}.
Any character sequence separated by whitespace can make up the events in the voices,
as long as they start vertically aligned to a duration symbol.

The vertical positions of the original text are reflected by assigning them to
the different
\tscore{} voices. The meaning of each such event is to hold down a particular
string in a particular fret and to pluck this string.
The encoding is given as part of the source text. This is indeed part of the
meta-model, not of the model.
Thus it can be given once in the general preamble of the file
(Figure~\ref{fig-newsidler} lines~6\textit{pp.}) and used
later in several parts. (line~15).

An additional suffix may indicate \emph{laissez vibrer}, that the string shall
sound longer then
the notated duration of its event. \citep[p.\,41]{wolf}

Since \tscore{} requires one duration symbol at each position of the \xxt T  line,
their typical jumps in the original documents
(see the top of Figure~\ref{fig-schlick}) is not taken over into the
\tscore{} source format. Instead, the prelude parameter \xxt{duratioCadens} says
that
the duration symbol appears always at the lowest possible graphical position.
This is reflected in the data output (by the attribute \xxt{\@ypos}, see below),
as well as by the generated control graphic, see the Figure.

The further parameter \xxt{duratioManet} allows in the time line
to use the carry operator \xxt-  which
stands for an empty space in the original, meaning that the previous
duration is still valid, see line~21 of Figure~\ref{fig-schlick}.

Line~20 of Figure~\ref{fig-newsidler} demonstrates the extensibility of
any \tscore{} meta-model:
Source lines for additional parameters can easily be added, for instance
editorial remarks in an event parameter with name ``\texttt{edit}''.

\xumbi{\newpage}
\section{Processing and Output Format}

The current implementation parses the input and generates as output an XML
data model.
This has a hybrid design which mixes graphical and semantical properties freely,
only guided by practical needs and thus only intermediate.
Transformation into more sensible data formats like
a specially taylored MEI variant \citep{mei} can easily be done with standard tools.
Figure~\ref{fig-newsidler-xml} shows the beginning of the (notoriously verbose) XML
format text representation of the resulting intermediate model.

Beside this, a simple graphic rendering is generated, for control purpose,
as also shown in the Figures.

The document type of the intermediate data model is Figure~\ref{fig-dtd}.
It contains~\ldots
{
\begin{itemize}\setlength{\itemsep}{0.0pt}\setlength{\parsep}{0.0pt}\setlength{\parskip}{0.0pt}
\item one \xxt{tabulatura} element for each source file \xxt{PARS}
\item one \xxt{columna} element for each column in the source text
\item one \xxt{duration} element for each such column, with the XML
  attributes~\ldots
\makeatletter
  \begin{itemize}\setlength{\itemsep}{0.0pt}\setlength{\parsep}{0.0pt}\setlength{\parskip}{0.0pt}
  \item \xxt{@ypos} vertical position in the original source layout
  \item \xxt{@trabes} whether beams end or begin at the corresponding stem
  \item \xxt{@duration} the duration of this column, encoded by its ``modern''
    reading, as a rational number with natural number numerator and denominator.
  \item \xxt{@summaPraecendentium} the sum of all preceding durations, giving the
    temporal position of the column, also as a rational number
  \item (\xxt{@numerus} mere technical numbering for identification.)
  \end{itemize}
\item one \xxt{sonum} element for each grip in this column, with the XML
  attributes~\ldots
  \begin{itemize}\setlength{\itemsep}{0.0pt}\setlength{\parsep}{0.0pt}\setlength{\parskip}{0.0pt}
  \item \xxt{@fret} where to put the finger
  \item 
    \xxt{@string}
       which string to press and pluck
  \item \xxt{@prolongate} whether this event has a ``laissez vibrer'' suffix
  \item \xxt{@ypos} vertical position in the original source layout
  \item (\xxt{@finger} which finger to use, as said by the source.
    Not yet supported.)
  \end{itemize}

\end{itemize}
\makeatother
}

\section{Current State and Possible Extensions}

The current version is just a proof of concept.
The idea and the software is in the public domain (CC-BY-NC-SA). The publication
as open source is currently in progress.

For a broader applicability some extensions seem desirable:
\begin{itemize}\setlength{\itemsep}{0.0pt}\setlength{\parsep}{0.0pt}\setlength{\parskip}{0.0pt}
\item The letter which represents  \emph{laissez vibrer} in the source text
  should be selectable by a score prelude parameter.
\item The tuning of the strings can be included in the prelude parameters,
  and thus the
  intermediate data structure can contain pitch information.
\item Some authors include \emph{fingering indications}, which can be included
  by defining further suffixes. \citep[p.\,42]{wolf}.
\end{itemize}

Anyhow, the work invested brought forward the development of \tscore{} as
such: We learnt that also the entries in the \xxt T  lines sometimes need
full-fledged event syntax and semantics.

\bibliographystyle{apalike}
\bibliography{tablature}

\end{document}